\documentclass[useAMS,usenatbib]{mn2e}

\usepackage{threeparttable}
\usepackage{pdflscape}
\usepackage{graphicx}
\usepackage{longtable}
\usepackage{url}
\usepackage{mathtools}

\title[H$_{2}$O maser astrometry: S235AB-MIR]{A `Water Spout' Maser Jet in S235AB-MIR}
\author[R. A. Burns]{R. A. Burns$^{1}$\thanks{E-mail:
RossBurns88@googlemail.com}, H. Imai$^{1}$, T. Handa$^{1}$, T. Omodaka$^{1}$, A. Nakagawa$^{1}$, T. Nagayama $^{2}$, \newauthor and Y. Ueno$^{2}$\\
$^{1}$Graduate School of Science and Engineering, Kagoshima University, 1-21-35 K\^orimoto, Kagoshima 890-0065, Japan\\
$^{2}$Mizusawa VLBI Observatory, National Astronomical Observatory of Japan,
2-12 Hoshigaoka-cho, Mizusawa, Iwate 023-0861, Japan}
\begin{document}

\date{Accepted 1988 December 15. Received 1988 December 14; in original form 1988 October 11}

\pagerange{\pageref{firstpage}--\pageref{lastpage}} \pubyear{2002}

\maketitle

\label{firstpage}

\begin{abstract}
We report on annual parallax and proper motion observations of H$_{2}$O masers in S235AB-MIR, which is a massive young stellar object in the Perseus Arm. Using multi-epoch VLBI astrometry we measured a parallax of $\pi = 0.63 \pm 0.03$ mas, corresponding to a trigonometric distance of $D= 1.56^{+0.09}_{-0.08}$ kpc, and source proper motion of ($\mu_{\alpha}\cos\delta$, $\mu_{\delta}$) = ($0.79\pm0.12$, $-2.41\pm0.14$) mas yr$^{-1}$. 
Water masers trace a jet of diameter 15 au which exhibits a definite radial velocity gradient perpendicular to its axis. 3D maser kinematics were well modelled by a rotating cylinder with physical parameters: $v_{\mathrm{out}} = 45 \pm 2$ km s$^{-1}$, $v_{\mathrm{rot}}=22 \pm 3$ km s$^{-1}$, $i = 12 \pm 2 ^{\circ}$, which are the outflow velocity, tangential rotation velocity and line-of-sight inclination, respectively.
One maser feature exhibited steady acceleration which may be related to jet rotation.
During our 15 month VLBI programme there were three `maser burst' events caught in the act which were caused by an overlapping of masers along the line of sight.  
\end{abstract}


\section{Introduction}

The formation of massive stars has been one of the most exciting astronomical problems of the recent decade. Forming stars above 8 $M_{\odot}$ cannot be explained by simply upscaling low mass star formation theories since radiation pressure limits further accretion \citep{Palla93}. Of the mainstream theories it is emerging that disk aided accretion is a strong candidate for explaining the formation of stars up to at least $B$ type, as observational evidence of disks associated with such accreting stars is found \citep{Hirota14,Beltran04}. Such protostellar disks are thought to be able to shield accreting material from stellar radiation and transport it down to radii of $\sim$100 au from the central object \citep{Zinnecker07,Cesa07}.


However, one necessity of disc accretion theory which has yet to be properly confronted with observations concerns how angular momentum is carried away from the protostellar disk to enable accretion. Proposed methods of removing angular momentum include Alven waves in the disk material, magnetic breaking in the ionic component of the rotating disk, and by physically carrying rotating material away from the disk vertically via magnetocentrifugally driven rotating jets (\emph{see review by} \citealt{Konigl00}). 

In addition to disk observations it is therefore also appealing to look for evidence of rotating jets near massive young stellar objects (MYSOs) in order to uncover their role in the angular momentum budget of accreting stars. A demonstration of such an investigation is the \emph{Ori-S6} MYSO \citep{Zap10}.
This kind of work poses an observational challenge since MYSOs are deeply embedded and rarely have observable collimated jets \citep{Navarete15}. We therefore turn to masers in our investigation of rotation in MYSO jets.

Collisionally excited H$_{2}$O masers are associated with a diverse range of kinematic phenomena in the young stellar objects (YSOs) of star forming regions (SFRs), including jets, wide-angle outflows and expanding shells (\emph{see for example}~\citealt{Torr12}). The bright and compact masers are ideal for very long baseline interferometry (VLBI) observations, which can penetrate the obscuring parent cores in which massive stars form and reveal the motions of young jets \citep{Moscadelli11,Furuya00}. Multi-epoch observations allow determination of the internal 3D kinematics at shocked regions via combination of line-of-sight and proper motion measurement. Furthermore the annual parallax of a maser source provides a precise estimation of its distance, which is crucial for evaluating its physical properties. As such VLBI is an indispensable technique for studying the formation of stars and provides a mechanism for the search for rotating MYSO jets.



VLBI observations in this paper were carried out using VERA (VLBI exploration of radio astrometry) \citep{Koba03} which is a Japanese VLBI network dedicated to maser astrometry of primarily Galactic sources. The bright H$_{2}$O maser in S235AB was chosen as a VERA target for its interesting `maser burst' behaviour (up to 120 Jy) \citep{Felli07} which has a dominant blue-shift maser (DBSM) spectrum. Its annual parallax was also sought for it's contribution to the mapping of the Perseus Arm.


In addition to H$_{2}$O masers, the S235AB region also exhibits class I methanol masers \citep{Krutz04,Pratap08} and two bipolar outflows of HCO$^{+}$, driven from a dusty molecular core seen in 450 and 850 $\mu m$ \citep{Felli04}. \citet{Felli06} proposed a \emph{class I} mid infrared source, S235AB-MIR, as the outflow progenitor from \emph{Spitzer} images. By SED modelling of sub-millimeter continuum data from \citet{Felli04} and \emph{Spitzer} mid-infrared photometries, \citet{Dew11} showed S235AB-MIR to be the only massive YSO in the S235AB cluster, with a stellar mass of 11 $M_{\odot}$ and still accreting. 

No centimeter emission has been detected in S235AB-MIR \citep{Tofani95,Felli06}, suggesting no significant development of an ultra compact H$_{\rm II}$ region (UCHIIR). \citet{Saito07} estimated the most massive member of the S235AB cluster to be of spectral class B1 from C$^{18}$O observations. Thus, a plausible interpretation above picture is that of a deeply embedded accreting $B$-type star which is too young to produce a detectable UCHIIR.


\vspace{-0.2cm}
\section{Observations and Data Reduction}

Simultaneous VERA observations of S235AB-MIR and the continuum reference source J0533+3451 were made in dual-beam mode \citep{dual}. Observing the maser and reference source simultaneously removes the need to slew antennae and interpolate phase measurements between sources, enabling excellent determination of the dynamic troposphere phase contribution (\emph{see} \citealt{Asaki07}). 

Phase tracking centers were set to
$(\alpha, \delta)_{\mathrm{J}2000.0}=(05^{\mathrm{h}}40^{\mathrm{m}}53^{\mathrm{s}}.38445496$,
+35$^{\circ}$41'48".4470) and
$(\alpha, \delta)_{\mathrm{J}2000.0}=(05^{\mathrm{h}}33^{\mathrm{m}}12^{\mathrm{s}}.7651060$,
+34$^{\circ}$51'30".336990) for S235AB-MIR and J0533+3451, respectively.
The continuum source J0533+3451 is listed in the VLBA calibrator list \citep{Petrov12} and was observed at an unresolved $K$-band flux of $\sim$20 mJy in our observations with VERA.
Intermittent observations of BL Lac, DA55 or 3C84 were made every 1.5 hrs for bandpass and group delay calibration.
Typical observing sessions were 8 hrs long, providing 2.5 hrs on-source time and sufficient \emph{uv}-coverage.

Left-handed circular polarisation signals were sampled at 2-bit quantisation, and filtered with the VERA digital filter unit \citep{Iguchi05}.
The total bandwidth of 256 MHz was divided into 16 intermediate frequency (IF) channels, each with a bandwidth of 16 MHz. One IF was allocated to the maser signal, assuming a rest frequency of 22.235080 GHz. The other 15 IFs, in adjoining frequency, were allocated to J0533+3451. Signal correlation was carried out using the Mitaka FX correlator \citep{Chikada}. Channel spacings of 31.25 kHz were used for the maser data, corresponding to a velocity resolution of 0.42 km s$^{-1}$.

\begin{table}
\caption{Summary of observations made with VERA.\label{obs}}
\begin{center}
\small
\begin{tabular}{cccccc}
\hline
Observation&&&Detected\\
Epoch&Date&MJD&spots\\ \hline
1&2013 Jan 29&56321&17\\
2&2013 Feb 21&56344&18\\
3&2013 Mar 15&56366&26\\
4&2013 Apr 21&56403&12\\
5&2013 Sep 29&56564&30\\
6&2013 Nov 02&56598&24\\
7&2013 Dec 02&56628&26\\
8&2014 Jan 28&56685&21\\
$^{\dag}$9&2014 Mar 11&56727&19\\
$^{\dag}$10&2014 Apr 29&56776&19\\
\hline
\end{tabular}
\begin{tablenotes}
\item $\dag$ These epochs were not used in parallax determination.
\end{tablenotes}
\end{center}
\end{table}

Global positioning system (GPS) measurements of the atmospheric water vapour content at each station were used to refine an a-priori model of expected atmospheric delay \citep{Honma08b}. Solutions were applied post-correlation. 

VERA observations were conducted in 10 epochs, closely spaced to support maser identification in view of the extreme variability of emission. The observation calendar is summarised in Table~\ref{obs}. An error (possibly caused by a shift in the adopted antenna positions) in our correlation settings in epochs 9 and 10 resulted in inaccurate astrometry such that data from these epochs could not be used in measurement of the annual parallax. These data were only used in maser distribution and line of sight velocity analysis.

All data were reduced using the Astronomical Image Processing System (AIPS) developed by the National Radio Astronomy Observatory (NRAO).

Data were reduced using the \emph{inverse phase-referencing} method for VERA data, which was introduced in \citet{Imai12}. In this approach the bright maser emission is used to calibrate dynamic tropospheric fluctuations in phase and rate - thus acting as the phase reference - while group delay is solved using a bright continuum source. These solutions are applied to the data of a nearby continuum source whose precise co-ordinates are already known. The relative position of the maser to the nearby continuum source gives the accurate position of the maser.
A detailed explanation, a flowchart and an internet link to a data reduction walkthrough are given in the Appendix of this paper.

The main advantage of the inverse phase-referencing technique is its ability to detect weak continuum sources ($\sim$10 mJy, \citealt{Imai12}). Furthermore, since the phase-rate solutions applied to the nearby reference source are also used to self-calibrate the maser data, the phase-referenced maser maps have high dynamic range.

Emission peaks in the maser maps were imaged using the CLEAN procedure, based on \citet{Hog74} and detections were registered at a signal-to-noise cutoff of 5. Following common nomenclature, a maser `spot' refers to an individual maser brightness peak, imaged in a specific velocity channel, and a maser `feature' refers to a group of spots which are considered to emanate from the same physical maser cloud. Maser spots are categorised into features when they are grouped within 1 mas of another spot, and are continuous in velocity. Features and their associated spots are denoted by letters and numbers respectively in column 1 of Table~\ref{TAB}.


\section{Results}

\subsection{Maser detections}

In total, 19 maser features were catalogued comprising some $\sim$60 spots, all of which showed notable time variability. Maser detections are summarised in Table~\ref{TAB}.
Features can be separated into three main groups by comparing local standard of rest (LSR) velocities to that of the molecular core, which has a velocity of $-17$ km s$^{-1}$ \citep{Felli04}. The groups are: `red lobe', `core velocity' and `blue lobe' masers, which were observed from $+2$ to $+14$, around $-18$, and from $-55$ to $-65$ km s$^{-1}$, respectively.

Red lobe and core velocity masers were detected sporadically, and were weak - with typical fluxes of $\leq1$ Jy. The red lobe group had only two individual maser features and the core velocity masers had only one maser feature. Masers from these groups reside loosely together in a region of about 120 $\times$ 120 au.

Blue lobe maser emission was strong, extremely variable and exhibited a `burst' nature where the cross-power flux rose from $10^{0}$ Jy to $10^{2}$ Jy on three occasions. These masers are blueshifted $>40$ km s$^{-1}$ with respect to the molecular core. Blue lobe masers reside in a compact region of about 15 $\times$ 15 au at a separation of 2" from the other maser groups.

\begin{figure}
\begin{center}
\hspace{-0.6cm}
\includegraphics[scale=0.76]{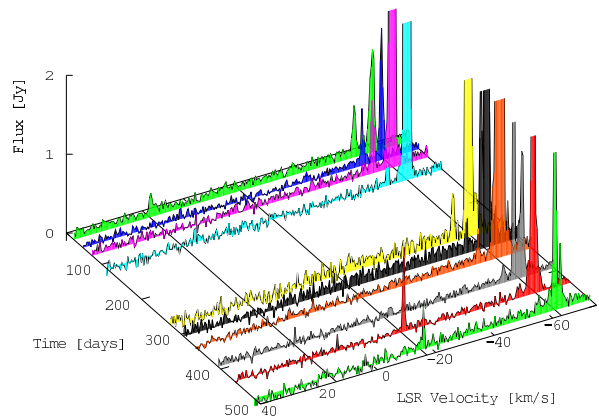}
\caption{Scalar averaged spectra of the H$_{2}$O emission for each epoch, limited to 2 Jy to show weaker detections. Lines of constant velocity indicate the red lobe, the core velocity and the blue lobe maser velocities in descending velocity order.
\label{SPEC}}
\end{center}
\end{figure}

Relative intensities and time variation of the aforementioned velocity components are shown in Figure~\ref{SPEC}. The spectrum of S235AB-MIR exhibits a classical DBSM profile, providing early evidence that the water masers may be driven by a jet \citep{CP08,Motogi13,Motogi15}.
Figure~\ref{ALL} shows the velocities and distributions of all detected maser features, revealing a jet. A dashed line drawn through the central positions of the main maser groups traces a position angle of $PA = 217^{\circ}$ which delineates the geometric axis of the system. The right-hand panel of Figure~\ref{ALL} shows a position-velocity (p-v) diagram across this axis.

\subsection{Parallax and proper motions}
Astrometric motions of masers are a combination of a linear component corresponding to the proper motion of the maser on the sky, and a sinusoidal component due to annual parallax. These two components are separated by simultaneous fitting of a linear and sinusoidal function to the observed data. Long temporal baselines and dense sampling are vital to correctly separate the two components, therefore we only fit parallaxes using masers associated with the persistent features; A and B.

We performed data fitting using 3 different approaches: individual, group and feature fitting. 
`Individual fitting' refers to fitting the parallax and proper motion of each individual maser spot independently. Fitting is done using the same velocity channel at each epoch i.e. we assume non-acceleration in the line of sight. We use only the most stable masers that were identifiable in at least 5 epochs and persist for at least 1 year. 

`Group fitting' involves fitting the astrometry of multiple spots together by assuming a common distance. In this procedure we use all masers that are identifiable in at least 4 epochs and are associated with features A or B. We perform `group fitting' on masers in features A and B together, and also separately for comparison.

`Feature fitting' is an approach where we determine the nominal position of the entire maser feature by calculating the flux-weighted average position of the three brightest maser spots. 

The results of these three fitting procedures are summarised in Table~\ref{TAB}, where astrometric positions in columns 4 and 5 are quoted as offsets from the phase tracking center of the maser beam which was set to $(\alpha, \delta)_{\mathrm{J}2000.0}=(05^{\mathrm{h}}40^{\mathrm{m}}53^{\mathrm{s}}.38445496$,
+35$^{\circ}$41'48".4470). 

`Individual fitting' was done on 3 masers in feature A. Each maser spot gave consistent parallax and proper motion values. The `group fitting' of maser spots in feature A gave results that were consistent with the average of the `individual fitting' results. However, when using `feature fitting' the same consistency could not be obtained. 
For thermal sources the flux of each component should be steady, thus the nominal position of the feature should trace the gravitational center. However, feature A exhibits multiple velocity components as can be inferred from column 3 of Table~\ref{TAB} (see table footnote) and any changes in the relative brightnesses of components will affect the determination of the nominal astrometric position of the feature. Therefore, regarding feature A, fitting is best done using spots rather than features. 

Multiple maser components also exist in feature B as is confirmed by its maser burst nature (\emph{see subsection 4.2}). As such, `feature fitting' was also inappropriate for feature B. Emission was not detected for a whole year in any one channel barring `individual fitting', however `group fitting' for this feature gave a parallax that matched very well to the results for feature A.

Maser emission in feature C per individual epoch is narrower in velocity than that of features A and B, and does not exhibit multiple velocity peaks or a burst nature. This suggests that feature C has a simpler physical and velocity structure than features A and B. However, feature C is seen to increase in velocity over time (\emph{see subsection 4.5}). As a result, the temporal baseline of emission seen in any one maser channel was too short for astrometry using `individual fitting' or `group fitting'. The proper motion of this maser feature was obtained through `feature fitting', its parallax value was consistent with the aforementioned results from features A and B, however the parallax was not used as it violates the criterion that masers must not be accelerating.


In order to conclude the annual parallax of S235AB-MIR we decided to use `group fitting' for maser spots in features A and B together. 
Error floors (\emph{see} \citealt{Hachi09}) of 0.19 and 0.15 mas were added to RA. and Dec. positions, respectively, in order to achieve a $\chi^{2}=1$ fit.
The annual parallax of S235AB-MIR was $\pi = 0.639 \pm 0.033$ mas corresponding to a trigonometric distance of D $= 1.56^{+0.09}_{-0.08}$ kpc, firmly placing it in the Perseus Arm.

The failure of `feature fitting' for features A and B demonstrates that the type of fitting procedure used can affect parallax estimates, especially in the presence of unresolved structure. Proper motions shown in Table~\ref{TAB} are those determined in by group fitting, assuming this common distance. Maser astrometries and their components are shown in Figure~\ref{squiggle}.

\null

\subsection{Internal and systemic motions}

The observed proper motions of Galactic masers are a combination of internal motions inherent in the SFR, and the systemic motion of the SFR with respect to the Sun.



Considering Galactic motion, S235AB-MIR resides in the Galactic anticenter direction ($l = 173.7^{\circ}$), where systemic proper motions are small. 
Using the Galactic rotation curve from \citet{Reid14}:
$\Theta = \Theta_{0} - 0.2(R-R_{0})$, where $R_{0}=8.34 \pm 0.16$ kpc, $\Theta_{0}=240 \pm 8$ km s$^{-1}$, and using our distance to S235AB-MIR; 1.56 kpc,
we calculate $\Theta = 240$ km s$^{-1}$. From this, we estimate the proper motion contribution due to Galactic rotation using a formulation from \citet{Ando11}:\\

$\mu_{l} cos b = \frac{1}{a_{0}D} \left[ \Big( \frac{\Theta}{R} -\frac{\Theta_{0}}{R_{0}} \Big) R_{0} cos~l - \frac{\Theta}{R} D\right] $\\

\noindent Which gives $\mu_{l} cos b = -0.019$ mas yr$^{-1}$. We must also correct for the motion of the Sun with respect to the LSR, which is taken as $(U, V, W)_{\odot} = (+10.3, +15.3, +7.7)$ km $s^{-1}$ from \citet{Ando11} (based on \citealt{Kerr86}). These contribute an apparent motion of $(\mu_{\alpha} cos\delta , \mu_{\delta}) = (+0.27, -2.41)$ mas yr$^{-1}$. By subtracting this motion from the observed proper motions we isolated the internal motions of water masers in S235AB-MIR which are shown in Figure~\ref{ALL}. 

Alternatively we sought to infer the internal proper motions by use of a reference maser. Feature F has a velocity that is well removed from those of the blue and red lobe masers and instead has a velocity close to that of the molecular core; this feature seems unrelated to the jet.
By designating feature F as the reference maser we can separate the internal motions (traced by blue and red lobe masers) from the source systemic motion (traced by feature F).

Feature F has a proper motion of $(\mu_{\alpha} cos\delta , \mu_{\delta}) = (+0.79 \pm 0.12, -2.41 \pm 0.14)$ mas yr$^{-1}$ which is almost identical to the correction factor calculated above from the LSR-frame motion. As such both the reference maser approach and LSR-frame approach produce very similar internal motions. In the latter approach the source systemic proper motion would be equivalent to that of feature F; $(\mu_{\alpha} cos\delta , \mu_{\delta}) = (+0.79 \pm 0.12, -2.41 \pm 0.14)$ mas yr$^{-1}$. This value is consistent with other Perseus Arm sources catalogued in \citet{Reid14}.

Resultant internal proper motions are typically smaller than the line of sight velocities and motions do not appear well collimated. This can be explained by projection effects from a pole-on geometry, as was seen in the case of G353.273+0.641 which is also a DBSM jet source \citep{Motogi15}.

\section{Discussion}

\subsection{Driving source: S235AB-MIR}

\citet{Felli06} proposed the \emph{class I} MYSO S235AB-MIR as the progenitor of the molecular outflows and maser emission in this region. Our observations confirm that the blue-shifted and red-shifted velocity components, which are separated by 2", lie either side of S235AB-MIR and in agreement with the corresponding blue and red-shifted lobes of the NNW-SSE molecular outflows seen in HCO$^{+}$ \citep{Felli04}. Less extended blue and red-shifted lobes, also orientated NNW-SSE, are seen in C$^{34}$S. \citet{Felli06} consider two possible interpretations of the C$^{34}$S velocity lobes; a molecular outflow and a rotating disk. 
Positions, line of sight velocities, and proper motions of water masers presented in this work support the outflow interpretation of \citet{Felli06}, such that the C$^{34}$S gas is a component of the NNW-SSE outflow. Furthermore, the proper motions of water masers oppose the radio and millimeter source VLA-1/M1 as progenitor of the NNW-SSE outflow, rather supporting the view that the outflows are driven by S235AB-MIR.




The HCO$^+$ emission has a distinct quadrupole morphology - one bipolar set of lobes orientates to NNW-SSE with the other set almost perpendicular. At the current best resolution the outflows appear concentric. Regarding establishment, one possible explanation is that of a sudden change in the outflow direction of a single bipolar outflow driven by the MYSO in S235AB-MIR, due to an encounter with another cluster member \citep{Bally05}. Another possibility is that of multiple bipolar outflow progenitors in the S235AB-MIR core, at angular sizes finer than those probed in the \emph{Spitzer} images. Considering this scenario, we cannot rule out the possibility that the high velocity water masers in S235AB-MIR are driven by a companion low mass YSO. However, the emission maps of \citet{Felli04} show that the orientation and velocity structure of the C$^{34}$S (dense gas tracer) emission correlates with that of the NNW-SSE HCO$^{+}$ outflow and water masers (this work), whereas no C$^{34}$S gas was seen to trace the NE-SW outflow. So, it is reasonable to assume that the former may therefore be energetically dominant and would thus likely be attributed to the MYSO.

\onecolumn


\begin{table}
\scriptsize
\vspace{-0.2cm}
\begin{center}
\caption{The general properties of H$_{2}$O masers in S235AB-MIR detected with VERA. \label{TAB}}
\begin{tabular}{cclcccccccc}
\hline
Maser&$V_{\rm LSR}$&Detected &$\Delta \alpha \cos \delta$&$\Delta \delta$ &$\mu_{\alpha}\cos\delta$&$\mu_{\delta}$&$\pi$\\ 
ID &(km s$^{-1}$)&epochs&(mas)&(mas)&(mas yr$^{-1}$)&(mas yr$^{-1}$)&(mas)\\
\hline
 $A1$  & $-60.06$  &**345**8** & $-6.95$&$32.45$  &  $ 0.62\pm0.27$& $-1.95\pm0.22$  &\\ 
 $A2$  & $-60.48$  &**3*5**8** & $-6.92$&$32.47$               &                &                 &\\ 
 $A3$  & $-60.90$  &1*3\textbf{\underline{4}}56*** 10 & $-6.76$&$32.88$      &  $ 1.30\pm0.29$& $-2.36\pm0.23$  \\ 
 $A4$  & $-61.32$  &1234\textbf{\underline{56}}789 10 &$-6.72$&$32.88$ &  $ 1.17\pm0.18$& $-2.42\pm0.15$   &$0.713\pm 0.092$\\ 
 $A5$  & $-61.74$  &1\textbf{\underline{23}}*56\textbf{\underline{7}}8* 10   &$-6.68$&$32.91$  &  $ 1.34\pm0.19$& $-2.48\pm0.15$  &$0.675\pm 0.061$\\ 
 $A6$  & $-62.16$  &1***567\textbf{\underline{8}}9 10   &$-6.69$&$33.01$& $ 1.20\pm0.25$& $-2.43\pm0.20$ & $0.558\pm 0.083$&\\ 
 $A7$  & $-62.58$  &\textbf{\underline{12}}**56*89 10  &$-6.65$&$33.09$ &  $ 1.24\pm0.22$& $-2.43\pm0.18$  \\ 
 $A8$  & $-63.00$  &12**5**89 10           &$-6.61$&$33.09$ &  $ 1.08\pm0.23$& $-2.38\pm0.18$  \\ 
 $A9$  & $-63.42$  &1\textbf{\underline{2}}**5**8\textbf{\underline{9}} \textbf{\underline{10}}  &$-6.54$&$33.02$ &  $ 0.95\pm0.23$& $-2.37\pm0.18$  \\ 
 $A10$ & $-63.84$  &123*\textbf{\underline{5}}***9 10  &$-6.44$&$32.97$ &  $ 0.60\pm0.37$& $-2.36\pm0.29$  \\ 
 $A11$ & $-64.26$  &*23******* &$-6.57$&$32.81$\\ 
 $A12$ & $-64.68$  &**3******* &$-6.49$&$32.70$\\ 
 $A13$ & $-65.10$  &**\textbf{3}*******  &$-6.44$&$32.74$\\ 
 $A14$ & $-65.52$  &**3******* &$-6.43$&$32.76$\\ 
\hline
 &&&&&&Average&$\pi =0.649 \pm 0.144$\\ 
\hline
&Feature A group fitting &&&& $1.06\pm 0.08$&$-2.35 \pm 0.07$ & $ \pi = 0.656 \pm 0.038$\\
&Feature fitting result &&&&$0.36\pm0.19$&$-2.45\pm0.10$& $ \pi = 1.018 \pm 0.093$\\
\hline
\hline
 $B1$  & $-60.06$  &***4******&$-6.33$&$36.85$&                &                 \\ 
 $B2$  & $-60.48$  &***4*6****&$-6.27$&$36.81$&                &                 \\ 
 $B3$  & $-60.90$  &**34*67***&$-6.67$&$37.18$&  $1.26\pm0.31$& $1.47\pm0.24$    \\ 
 $B4$  & $-61.32$  &*2345\textbf{\underline{6}}7***&$-6.56$&$37.05$&  $1.34\pm0.26$& $1.10\pm0.20$    \\ 
 $B5$  & $-61.74$  &*\textbf{\underline{2}}3\textbf{\underline{4}}56\textbf{\underline{7}}***&$-6.53$&$37.05$&  $1.54\pm0.26$& $1.17\pm0.20$    \\ 
 $B6$  & $-62.16$  &\textbf{\underline{1}}23*567***&$-6.42$&$37.28$&  $1.43\pm0.23$& $0.86\pm0.18$    \\ 
 $B7$  & $-62.58$  &123*\textbf{\underline{5}}67***&$-6.47$&$37.14$&  $1.45\pm0.23$& $0.92\pm0.18$    \\ 
 $B8$  & $-63.00$  &*2\textbf{\underline{3}}*5678**&$-6.24$&$36.98$&  $1.52\pm0.22$& $0.68\pm0.18$    \\ 
 $B9$  & $-63.42$  &*23*5\textbf{\underline{6}}78**&$-6.15$&$36.86$&  $1.40\pm0.22$& $0.83\pm0.18$    \\ 
 $B10$ & $-63.84$  &**3*56\textbf{\underline{7}}\textbf{\underline{8}}**&$-6.14$&$37.02$&  $1.42\pm0.29$& $0.63\pm0.23$    \\ 
 $B11$ & $-64.26$  &**3*5678**&$-6.10$&$36.99$&  $1.43\pm0.29$& $0.64\pm0.23$    \\ 
 $B12$ & $-64.68$  &**3*\textbf{\underline{5}}67***&$-6.11$&$36.98$&  $1.31\pm0.35$& $0.88\pm0.28$    \\ 
 $B13$ & $-65.10$  &**3*567***&$-6.21$&$37.06$&  $2.17\pm0.35$& $0.64\pm0.28$    \\ 
 $B14$ & $-65.52$  &****56\textbf{\underline{7}}***&$-3.72$&$37.33$&              &                   \\ 
 $B15$ & $-65.94$  &****567***&$-3.99$&$37.20$&              &                   \\ 
 $B16$ & $-66.36$  &****5*7***&$-4.06$&$37.16$&              &                   \\

\hline
&Fearure B group fitting &&&& $1.48\pm 0.08$&$0.89 \pm 0.07$& $ \pi = 0.628 \pm 0.057$\\
&Feature fitting result &&&&$1.89\pm0.14$&$0.74\pm0.21$& $ \pi = 0.449 \pm 0.069$\\
\hline
\hline
&Collective group fitting of \\
&spots in features A and B &&&&&& $ \pi = 0.639 \pm 0.033$\\
\hline
\hline
$C1$ & $-59.22$  &1\textbf{\underline{23}}******* &$-8.103$&$33.71$\\ 
$C2$ & $-59.64$  &\textbf{\underline{1}}********* &$-8.063$&$33.65$\\ 
$C3$ & $-60.06$  &1*34****9*          &$-7.935$&$33.594$\\ 
$C4$ & $-60.48$  &***\textbf{\underline{4}}****9* &$-7.568$&$33.221$\\ 
$C5$ & $-60.90$  &****5***9*         &$-6.271$&$33.644$\\ 
$C6$ & $-61.32$  &****\textbf{\underline{5}}6789 10&$-6.206$&$33.644$\\ 
$C7$ & $-61.74$  &****5\textbf{\underline{6}}789 10 &$-6.291$&$33.709$\\ 
$C8$ & $-62.16$  &****5678\textbf{\underline{9}} 10         &$-6.316$&$33.714$\\ 
$C9$ & $-62.58$  &****567\textbf{\underline{8}}9 \textbf{\underline{10}} &$-6.061$&$33.509$\\ 
$C10$ & $-63.00$  &******\textbf{\underline{7}}89 10 &$-6.498$&$33.529$\\ 
$C11$ & $-63.42$  &******789 10 &$-6.493$&$33.514$\\ 
$C12$ & $-63.84$  &******789 10 &$-6.528$&$33.509$\\ 
$C13$ & $-64.68$  &*******8** &$-6.393$&$33.251$\\ 
\hline
&Feature fitting result &&&&$1.37\pm 0.19$&$-0.44\pm0.15$&$ \pi = 0.632 \pm 0.077$\\
\hline

\hline
 $D$  & $+14.12$ &12********    &$-1213$&$1563$&   &      \\ 
 $E$  & $+2.73$   &*********10   &$-1155$&$1599$&               &                 \\
 $F$  & $-17.91$  &*******89 10 &$-1136$&$1619$& $0.79 \pm 0.12$ & $-2.41 \pm 0.14$  \\

 $G$  & $-61.32$ &1234******   &$-8.19$&$36.35$& $1.47\pm 0.20$& $-1.91\pm 0.20$  \\ 
 $H$  & $-60.91$ &****5*78**   &$-7.19$&$36.29$& $0.84\pm 0.22$& $ 1.44\pm 0.30$  \\ 
 $I$  & $-65.94$ &****567***   &$-1.68$&$36.30$& $3.46\pm 0.20$& $-2.02\pm 0.21$  \\ 
 $J$  & $-55.43$ &123*******   &$-9.36$&$32.22$& $2.60\pm 0.18$& $-6.18\pm 0.20$  \\ 
 $K$  & $-56.28$ &123*5*****   &$-5.18$&$26.21$& $3.77\pm 0.18$& $-5.61\pm 0.20$  \\ 
 $L$  & $-56.70$ &***4******  &$-4.38$&$25.25$              &                  \\ 
 $M$  & $-57.54$ &********9*  &$-11.14$&$37.03$              &                  \\ 

 $N$  & $-61.32$ &****5*****  &$-1.41$&$30.14$              &                  \\ 
 $O$  & $-63.86$ &*********10 &$-3.12$&$29.39$              &                 \\ 
 $P$  & $-64.25$ &****5*****  &$-5.69$&$39.10$              &                  \\ 
 $Q$  & $-64.70$ &********9*  &$-5.06$&$33.18$              &                  \\ 
 $R$ & $-64.70$ &*********10 &$-1.67$&$35.90$&\\
 $S$ & $-64.70$ &*********10 &$-2.16$&$33.21$&\\
 
\hline
\end{tabular}
\begin{tablenotes}
\item{\footnotesize{Column (3)}: Epoch numbers indicate detection, while asterisk represents non-detection. For each epoch the brightest velocity component(s) in features A, B and C are indicated by underlined boldface digits.}
\end{tablenotes}
\end{center}
\end{table}

\begin{figure}
\begin{center}
\includegraphics[scale=0.92]{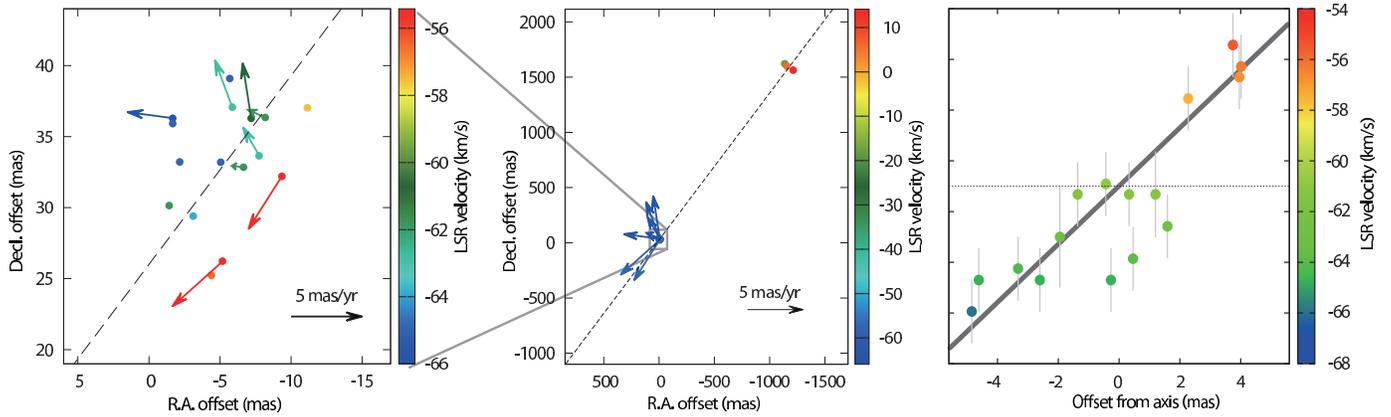}
\caption{Positions and internal motions (corrected for systemic motion) of masers in S235AB-MIR. Coordinates are given respective to the phase tracking center of the maser data. \emph{Middle} shows the full scale maser distributions including the three maser groups discussed in the main text. A short dashed line traces the jet axis. The proper motion of feature F ($v_{LSR}=-17.91$ km s$^{-1}$) is too small to see. \emph{Left} shows a zoom of the cluster of blue lobe masers. \emph{Right} shows a p-v diagram of the blue lobe masers with respect to the jet axis, where vertical bars are the half-widths of typical maser velocity features, larger bars at offsets of +1 and -2 mas correspond to features A and B, respectively, which have wider velocity profiles. The thick grey line is the least squares fit and traces a gradient of 1.2 km s$^{-1}$ mas$^{-1}$.
\label{ALL}}
\end{center}
\end{figure}

\begin{figure}
\vspace{2cm}
\begin{center}
\includegraphics[scale=1.6]{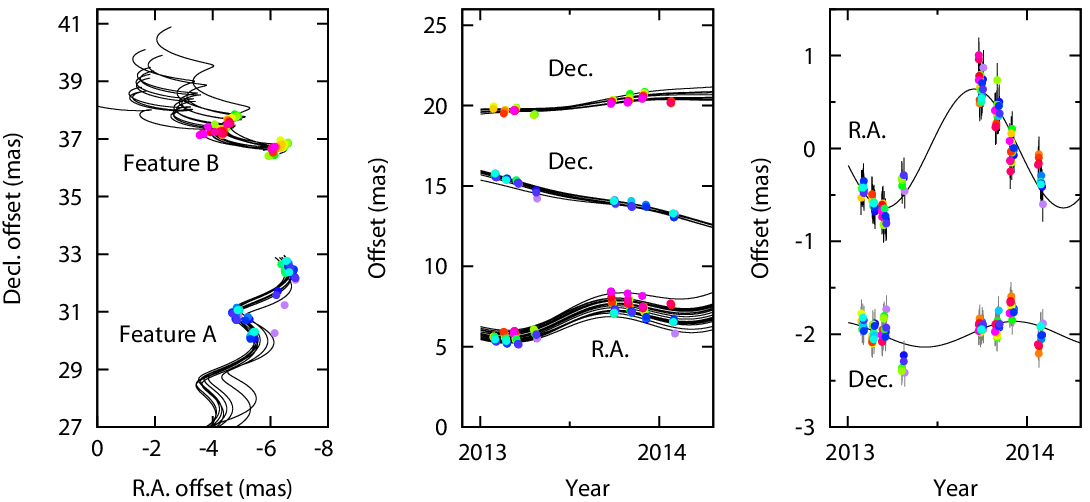}
\caption{Parallax and proper motion fitting of spots (coloured arbitrarily) associated with maser features A and B in S235AB-MIR. \emph{Left}: Sky-plane motions of the maser spots with respect to the phase tracking center. \emph{Middle}: arbitrary scale astrometric offsets in Dec. (above and middle for Features B and A, respectively) and R.A. (below) as a function of time. \emph{Right}: parallactic motion in R.A. (above) and Dec. (below) for all spots in features A and B after subtraction of linear proper motions. Small horizontal offsets (within $\pm 7$ days) were introduced to data points in the middle and right-hand plots for readability. 
\label{squiggle}}
\end{center}
\end{figure}

\twocolumn

\newpage

\subsection{Maser burst}
\begin{figure}
\begin{center}
\hspace{-1cm}
\includegraphics[scale=1.4]{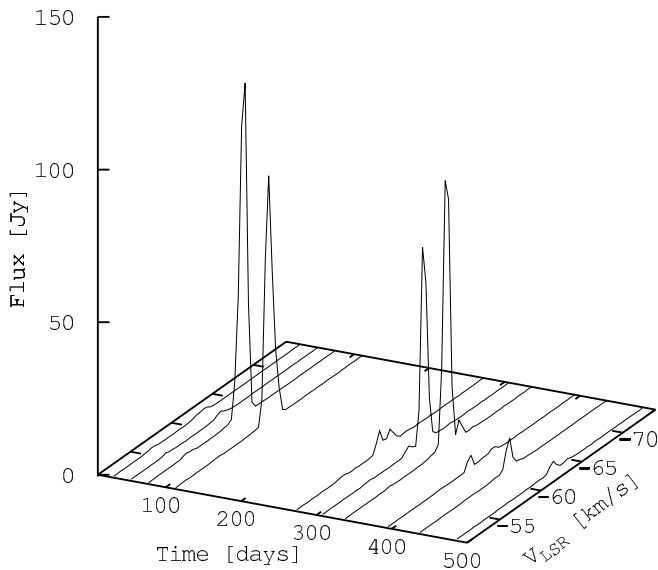}
\caption{Scalar averaged spectra of the blue lobe H$_{2}$O emission at full scale, highlighting the variable burst nature.
\label{SPEC2}}
\end{center}
\end{figure}

A sudden increase in maser brightness is known as a `burst'. \citet{DnW89} proposed a maser burst model in which an increase in path length, achieved by the overlapping of two masing clouds along the line of sight, can produce an increase of two or more orders of magnitude in observed brightness. Multi-epoch VLBI observations by \citet{Shimoikura05} found observational evidence of this process in the case of the water masers of Orion KL, however there are few other examples of a maser burst caught `in the act'. There were 2 strong maser bursts during our VLBI monitoring observations of S235AB-MIR in which fluxes of $\sim$100 Jy were reached, as shown in Figure~\ref{SPEC2}.

\begin{figure}
\begin{center}
\hspace{-1cm}
\includegraphics[scale=0.59]{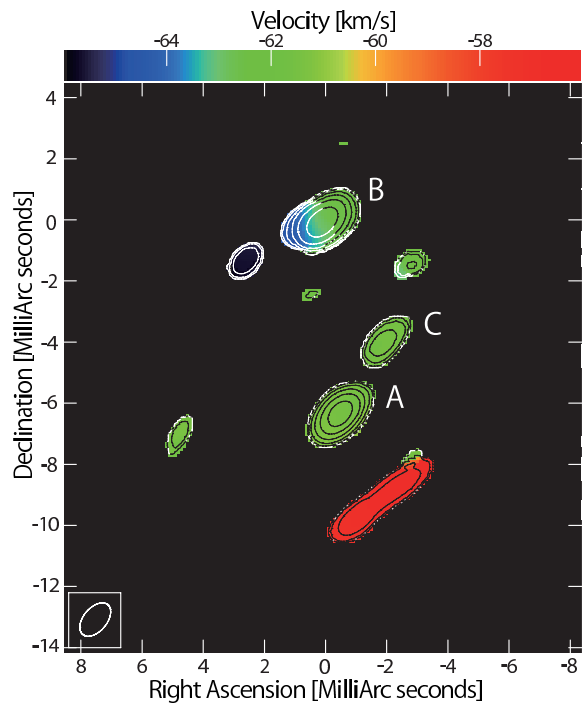}
\caption{First moment (velocity) map of water maser emission in epoch 5 (MJD 56564) where positions are relative to the reference maser in feature B. The three main features are labelled. Two velocity components in feature B, at map position (0,0), are the progenitors to the maser burst seen in epoch 6 (reaching 80 Jy). Contours are 3, 5, 10, 20, 40, 80 times the image rms noise of  200 mJy.
\label{BURST}}
\end{center}
\end{figure}

Column 3 of Table~\ref{TAB} shows the detected velocity components (rows) for each epoch (columns), where the brightest peaks in in each epoch are shown as underlined boldface digits for spots associated with features A, B and C. From the peaks we can see that feature B has a complex velocity structure; velocity drifts and multiple peaks. Two peaks can be seen in epoch 5 (MJD 56564) at $-64.68$ km s$^{-1}$ and $-62.58$ km s$^{-1}$ which respectively correspond to the blue and green components seen in Figure~\ref{BURST}, where feature B is at map offset (0,0). In epoch 6 (34 days later) the total flux of feature B raised from 7 Jy to 80 Jy as the two components overlapped both in space and velocity. Therefore we observe that the bursting mechanism in the masers of S235AB-MIR \citep{Felli07} is the same as that which was seen in \citet{Shimoikura05}. Figure~\ref{BURST} illustrates the `pre-burst' stage of this mechanism.

Bursting activity was also seen in feature B at epoch 3 in which it reached a flux of 120 Jy. The bursting mechanism in epoch 3 was the same as described above.

Feature A also has a complex velocity structure. In this case feature A appears as a single peaked maser feature in epoch 1, which gradually separates in velocity into two defined peaks by epoch 3, and then recombines at epoch 5 - where a modest maser burst raised its flux to 20 Jy from its usual flux of 2$\sim$3 Jy.

The peak velocities of the overlapping maser components involved in the burst often differed by a few km s$^{-1}$, i.e. the burst occurred at the intermittent velocity between the peaks. This suggests the existence of a significant amount of population inverted gas at velocities away from the original peaks.
Such behaviour indicates that the physical maser cloud may have a spatially unresolved structure and possibly some internal motions since the bursting components were part of the same maser feature. Such scales cannot be investigated directly at the current angular resolution of terrestrial VLBI arrays, however the presence of such substructures has recently been observed in the results of space VLBI observations (RadioAstron newsletter: \#19 - released 6th March 2013, and \#27 - released 31st March 2015).

\subsection{H$_{2}$O masers: the jet and the velocity gradient}

H$_{2}$O masers can be collisionally excited on the surface of shock fronts created when a protostellar jet pushes into ambient gas (\emph{ex: conical jet surface} \citealt{Mosca00}, U\emph{-shaped `micro-jet'} \citealt{Furuya00}). The large velocity difference in blue ($-66$ km s$^{-1}$) and red lobe ($+14$ km s$^{-1}$) masers in S235AB-MIR indicate that these masers are likely tracing a jet. Elongated red masers parallel to the jet axis, shown in Figure~\ref{BURST}, may represent shocked gas on the inner surface of the jet.

The jet axis determined by H$_{2}$O masers in Figure~\ref{ALL} is in good agreement with the orientations of large scale HCO$^{+}$ and smaller scale C$^{34}$S outflows from \citet{Felli04}. We made a p-v diagram using offsets from the jet axis plotted against line-of-sight velocity in the right-hand panel of Figure~\ref{ALL}. Maser features are well fit to a straight line as would be expected from solid body rotation. Its gradient of 1.2 km s$^{-1}$ mas$^{-1}$ corresponds to 0.77 km s$^{-1}$ au$^{-1}$ at 1.56 kpc.



Velocity gradients across jet axes have been reported in several YSO observations and simulations \citep{Zap10,Zap15,Coff04,Wotias05,Bacciotti02,Cerq06}. To explain these velocity structures there is current debate between supporters of the jet rotation interpretation \citep{Zap10,Zap15,Coff04,Wotias05,Bacciotti02} and those who propose alternative explanations \citep{Soker05,Cerq06}. Although we believe S235AB-MIR possibly evidences a rotating jet we consider other commonly suggested explanations of velocity gradients in jets, before discussing a model of jet rotation.


\subsubsection{Interpreting the velocity gradient}

Precessing jets are occasionally found in MYSOs \citep{Garatti15}. In this scenario the jet is comprised of a continuous ballistic ejection whose direction is time-dependent, causing the well-known `wiggle' morphology.
\citet{Cerq06} show that the velocity field created by a precessing jet and a rotating jet are difficult to distinguish in radial velocity profiles of emission line observations. Such conclusions are based on large scale gas motions encompassed within the precession angle; scales larger than the radius of the jet. The H$_{2}$O masers in S235AB-MIR trace scales comparable to the physical radius of the jet ($\sim$20 au, \citealt{Furuya00,Coff04}). A velocity gradient on a jet radius scale can only be formed if there is motion within the jet itself - thus rejecting a precession scenario.

Velocity structures may also arise from dispersion at the tip of a jet driven bow shock. In this scenario the shocked gas at the tip of the jet is forced edgeways as it decelerates into ambient material. \citet{Lee00} modelled jet-driven bow shocks and found their p-v diagrams to be symmetric across the central axis (their Figure 26). This contrasts with the p-v diagram of water masers in S235AB-MIR which is distinctly asymmetric (Figure~\ref{ALL}). Furthermore, radial velocity gradients seen in the bow shocks of \citet{Lee00} are several orders of magnitude lower than that observed in S235AB-MIR (roughly $2\times 10^{-3}$ km s$^{-1}$ au$^{-1}$ for the case of HH 111 in their Figure 27). The bow shock interpretation is therefore inconsistent with the velocity gradient seen in S235AB-MIR

If masers were produced in the shock front of an expanding arc it would be possible to reproduce a velocity gradient across the direction of collective motion if the arc were flattened on a plane, inclined to the observer and fanning out at constant velocity from a common driving source. However a fanned out jet scenario is a strictly ballistic model, requiring the proper motions of maser features to interpolate to a common kinematic center. In contrast, blue lobe masers in S235AB-MIR form a cluster, as opposed to an arc, and proper motions do not interpolate; some of the maser features in S235AB-MIR have proper motions that are parallel to the jet axis while some are near perpendicular to it - which is difficult to explain by purely ballistic motion.


We also note that blue lobe masers have been known since \citet{Henkel86} yet the maser group remains compact despite differences in line of sight velocities and divergent proper motions. This is difficult to reconcile using ballistic arguments since masers would be expected to have dispersed more by the present day.


Since only a modest number of maser features were detected in S235AB-MIR we cannot fully dismiss all ballistic interpretations of the observed velocity gradient using the current data alone and require confirmation from observations that sample jet gas directly. 
Nevertheless, without a satisfactory ballistic explanation for the kinematics of water masers in S235AB-MIR we proceed with a rotating jet as our working hypotheses. As we will see, a rotating jet interpretation forms consistencies with the aforementioned arguments that had conflicted with ballistic explanations. To our present knowledge S235AB-MIR is the first candidate rotating jet traced by water masers and, if confirmed with further observations, presents an ideal laboratory to study magnetocentrifugal jets in a \emph{class I} YSO with an accurate distance.

We created a kinematic model of a rotating cylinder to compare with the sky-plane distributions and 3D (sky-plane and line of sight) velocities of the water masers in S235AB-MIR. A cylindrical surface model was chosen for its analogy to an interface between a jet and the ambient gas - where masers are excited. Also the choice of a cylinder over a conical surface circumvents the ambiguity of the driving source position (which should not matter since jets are able to de-collimate and re-collimate). The fixed physical parameters of the model were the jet position angle, $PA$, and radius, $r$, which were readily determined from the maser maps (Figure~\ref{ALL}) as $PA = 217^{\circ}$, $r = 7.5$ au. The free parameters were the linear jet velocity, $v_{\mathrm{out}}$, the tangential rotation velocity, $v_{\mathrm{rot}}$, and the inclination to the line of sight, $i$. For comparison with observational data the modelled 3D motions were projected into the sky-plane and line of sight.


\begin{figure}
\begin{center}
\includegraphics[scale=0.47]{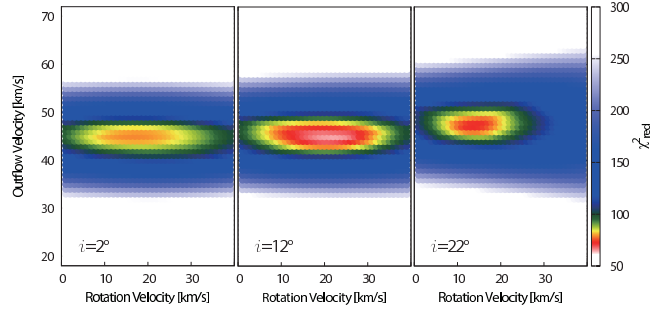}
\caption{Plots of $\chi^{2}_{\mathrm{red}}$ for fitting parameters at inclination slices of $i=2,12,22^{\circ}$ using model fitting condition 1.
\label{CHI}}
\end{center}
\end{figure}

Modelled values were compared with the maser data with regards to: positional offset from the jet axis, line-of-sight velocity with respect to the parent cloud ($-17$ km s$^{-1}$) and, when present, R.A and Dec. internal proper motions. We employed a `reduced $\chi^{2}$' fitting approach in which $\chi^{2}_{\mathrm{red}} = \chi^{2}/N$, where $N$ is the number of degrees of freedom, evaluated as $N = $ \emph{number of data points} $ - $ \emph{number of fitted parameters} $ - 1$. Variables that produce the lowest $\chi^{2}_{\mathrm{red}}$ correspond to the best match between model and data. For a check of consistency and to investigate the effects of varying degrees of freedom on $\chi^{2}_{\mathrm{red}}$ we considered 3 different conditions when modelling data, which were - to use data from: 1) all maser features. 2) only masers features with measured proper motions. 3) only maser features without measured proper motions (only position and line of sight velocities).

The resulting parameters were:\\
1) $v_{\mathrm{out}} =45$ km s$^{-1}$, $v_{\mathrm{rot}}=22$ km s$^{-1}$, $i = 12^{\circ}$: $\chi^{2}_{\mathrm{red}}=65$,
2) $v_{\mathrm{out}} = 45$ km s$^{-1}$, $v_{\mathrm{rot}}=19$ km s$^{-1}$, $i = 10^{\circ}$: $\chi^{2}_{\mathrm{red}}=55$,
3) $v_{\mathrm{out}} = 45$ km s$^{-1}$, $v_{\mathrm{rot}}=28$ km s$^{-1}$, $i = 10^{\circ}$: $\chi^{2}_{\mathrm{red}}=9$.


    


All fitting conditions generally agree on a high outflow velocity and a shallow inclination angle, with the most deviation seen in determination of the rotation velocity. We adopt $v_{\mathrm{out}} = 45$ km s$^{-1}$, $v_{\mathrm{rot}}=22$ km s$^{-1}$, $i = 12^{\circ}$ from condition 1 results since the fit used all data.

\begin{figure}
\begin{center}
\includegraphics[scale=0.9]{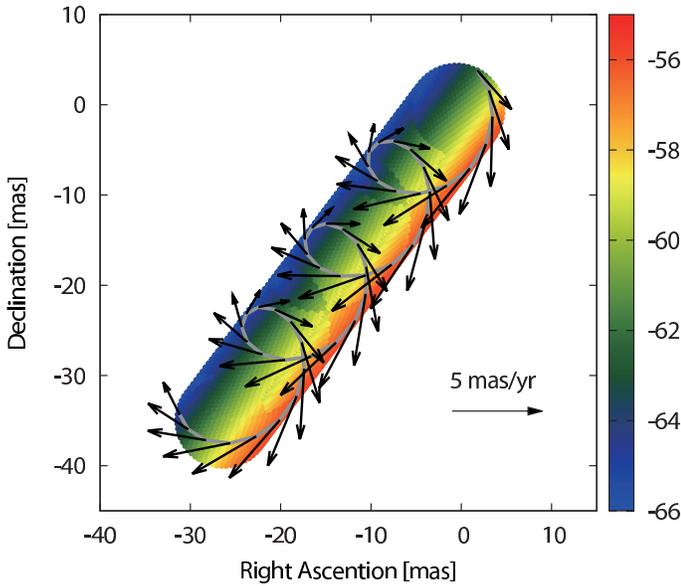}
\caption{Rotating cylinder model produced by the best fitting parameters. The grey line illustrates a test particle  trajectory in which vectors show the direction and magnitude of proper motions and colour indicates the line of sight velocity.
\label{MODEL}}
\end{center}
\end{figure}
\begin{figure}
\begin{center}
\includegraphics[scale=0.79]{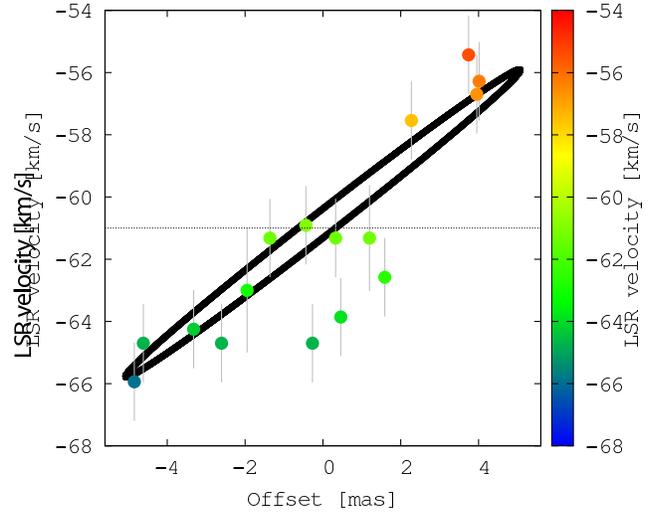}
\caption{Position-velocity diagram for maser features associated with the blue lobe of the jet where offsets are measured from the jet axis (from Figure~\ref{ALL}). Also shown is the position-velocity profile produced by our rotating cylinder model (\emph{solid line}). The horizontal line at -61 km s$^{-1}$ indicates the median velocity of the blue lobe masers.
\label{PV}}
\end{center}
\end{figure}

Values of $\chi^{2}_{\mathrm{red}}$ for parameters using condition 1 are shown in Figure~\ref{CHI} where the 3D ($v_{\mathrm{out}}, v_{\mathrm{rot}}, i$) distribution of values is represented in `slices' at inclination angles of $i=2,12,22^{\circ}$. Modelled and observed maser distributions and proper motions are well matched (Figures~\ref{ALL} \&~\ref{MODEL}) with the redder masers having larger proper motions in a more southernly direction and vice versa. Additionally the p-v diagram produced by the model shows good agreement with the p-v diagram drawn using the maser data (Figure~\ref{PV}).

\subsection{Overview of jet parameters}


To evaluate the success of the model fitting we calculated the scalar sum of 3D motions in the fitting residuals and compared this to the the scalar sum of 3D motions in the data. We find that the model can account for $80\%$ of observed motions, with $20\%$ remaining as residuals, which we adopt as the total model error.
The relative fitting success of each individual parameter ($v_{\mathrm{out}}$, $v_{\mathrm{rot}}$ and $i$) can be determined by the `3D' shape of the $\chi^{2}_{\mathrm{red}} $ locus in Figure~\ref{CHI}. In our fitting iterations $\chi^{2}_{\mathrm{red}}$ increases by $50\%$ its minimum value within $\Delta v_{\mathrm{out}} = \pm 4$,  $\Delta v_{\mathrm{rot}}= \pm 16$ and $\Delta i= \pm12$, corresponding to expected error contribution ratios of ($\epsilon_{v_{\mathrm{out}}}^{2}$:$\epsilon_{v_{\mathrm{rot}}}^{2}$:$\epsilon_{i}^{2}$) = (1$^{2}$:4$^{2}$:3$^{2}$).
The total model error can be separated into relative contributors, which allows us to assign error values to our best-fit model parameters, which become 
$v_{\mathrm{out}} = 45 \pm 2$ km s$^{-1}$, $v_{\mathrm{rot}}=22 \pm 3$ km s$^{-1}$, $i = 12 \pm 2 ^{\circ}$.




Our modelled rotation velocity of $v_{\mathrm{rot}}=22 \pm 3$ km s$^{-1}$ agrees well with those measured by \citet{Coff04} and \citet{Wotias05} on similar scales. The jet velocity gradient is linear and steep, at 1.2 km s$^{-1}$ mas$^{-1}$ which corresponds to 0.77 km s$^{-1}$ au$^{-1}$ at a distance of 1.56 kpc. The very steep gradient is consistent with the trend seen in \citet{Zap10} for velocity gradients to steepen at radii closer to the jet.

The outflow velocity component of the jet is comparable with those frequently reported in the literature and the inclination angle is typical for a DBSM source, which are also known to be jet driven \citep{CP08,Motogi13,Motogi15}. Regarding the age of the jet, if we take a symmetric case where the star stands between the red and blue lobes the jet length would be $r_{\star}= \mathrm{projected~distance} / \sin \, i = 1.5\times10^{4}$ au, using $D=1.56$ kpc. Therefore the approximate dynamical age of the maser jet is $t_{\mathrm{jet}} = r_{\star}/v_{\mathrm{out}} = 1600~years$, assuming constant velocity. This value will vary between $0<t_{\mathrm{jet}}<(2\times1600) ~\mathrm{years}$ depending on the position of the star. Note that this is not an estimate of the YSO age, which is $\sim 4000 ~\mathrm{years}$ \citep{Dew11}.

The low uncertainty seen in $v_{\mathrm{out}}$ is echoed by the consistency in the determination of $v_{\mathrm{out}}$ using fitting conditions 1, 2 and 3. This indicates that LSR velocity information dominates the determination of the outflow velocity. Larger uncertainty is seen in $v_{\mathrm{rot}}$ this parameter also shows less consistency when comparing the fitting conditions indicating that internal motions are important for characterising rotation. In the case that a better estimate for the systemic motion of the source becomes known we may refine the model of water masers in S235AB-MIR using more accurate corrections to internal motions.


Although ambiguities may remain in our model due to the small number of detected H$_{2}$O masers, it is important to note that \emph{all} models for all combinations of parameters and fitting conditions produce results that favour rotation in some way, \emph{consistently rejecting a ballistic-only interpretation}.

Evidence supporting the presence of a rotating jet is itself important as it promotes magnetocentrifugal jet motions as a viable mechanism for removing angular momentum from an accreting system. We do not quantify this angular momentum using our VLBI data because maser emission does not scale with mass. Masers have no net charge and thus their bulk motions are not directly influenced by magnetic fields. We propose that H$_{2}$O masers in S235AB-MIR are likely entrained in - and trace the kinematics of - a rotating jet of molecular gas. As such, we intend to revisit S235AB-MIR using millimeter array observations of jet tracers, which should provide suitable quantification of the angular momentum carried in the jet.


\subsection {Accelerating maser}

\begin{figure}
\begin{center}
\includegraphics[scale=0.72]{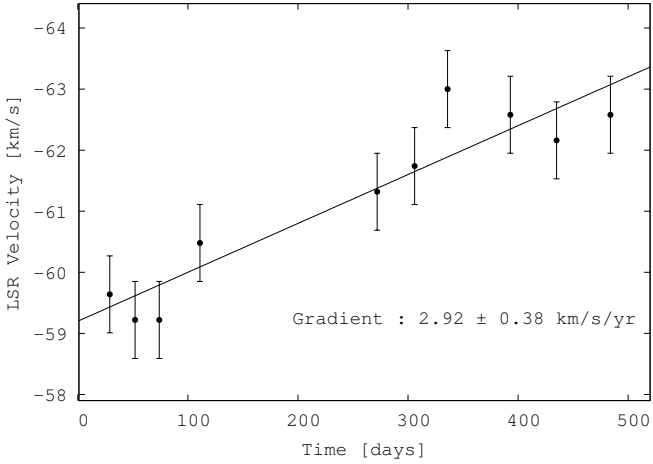}
\caption{Peak LSR velocity of feature C as a function of time, showing acceleration to bluer velocities.
\label{SALLY}}
\end{center}
\end{figure}

A stable velocity increase was observed in feature C, which was the only maser feature to exhibit such behaviour. We derived a velocity drift rate of $2.92 \pm0.38$ km s$^{-1}$ yr$^{-1}$, where the error value is the standard deviation of a linear fit to the data (Figure~\ref{SALLY}).

We introduced an additional error term of $\pm 0.63$ km s$^{-1}$ corresponding to the half-width of a typical maser spectral peak (these are the error bars in Figure~\ref{SALLY}) to account for the possibility that changes in velocity were misinterpreted from the changing in dominance of different velocity components in a maser feature. Subsequently the measured acceleration of feature C is $2.92 \pm1.01$ km s$^{-1}$ yr$^{-1}$.

Circular motions give rise to acceleration toward the axis of rotation therefore the appearance of an accelerating maser is a reasonable finding given the context. The measured acceleration is consistent with the model prediction, within the estimated uncertainty, which supports rotation as a possible origin and lends further support to the rotating jet interpretation of the velocity gradient in S235AB-MIR.


The velocity drift rate seen here is similar to that of DBSM maser source G353.273+0.641 \citep{Motogi15}. In their case, slower masers near to the driving source are accelerated up to the outflow velocity (120 km s$^{-1}$), thus the masers appear to trace launching motions at the root of the jet. In our case, however, the masers seem to be well removed from the driving source, supporting an alternative explanation.

\citet{Felli06} also observed a velocity drift in the water masers of S235AB using the single-dish Medicina radio telescope. They find two maser features in the $-5$ to 10 km s$^{-1}$ range which exhibit accelerations of $\sim 1$ km s$^{-1}$ yr$^{-1}$. The velocity range creates a near overlap between the core velocity and red lobe masers discussed in this work. \citet{Felli06} speculate that the velocity drift may have been due to the acceleration of masers from the central YSO up to faster redshift velocities (as seen in \citealt{Motogi15}). This is certainly feasible given the spatial consistency of these maser groups as seen in our VLBI images. Unfortunately these masers were rarely detected in our observations due to their weak emission. This possible maser outflow event would make an excellent target for VLBI observations at higher sensitivity.

\section{Conclusions}
Using VLBI monitoring observations we measured the astrometries of H$_{2}$O masers in S235AB-MIR. Our annual parallax of $\pi = 0.63 \pm 0.03$ mas corresponds to a distance of D $= 1.56^{+0.09}_{-0.08}$ kpc, placing the source in the Perseus Arm. The source systemic motion was $(\mu_{\alpha} cos\delta , \mu_{\delta}) = (+0.79 \pm 0.12, -2.41 \pm 0.14)$ mas yr$^{-1}$.

The S235AB-MIR molecular core drives a maser jet however it is unclear if the core contains a single MYSO or if it also houses companion lower mass YSOs. As such, the identification of the jet progenitor remains pivotal and dictates the significance of the VLBI maser data.

We present evidence that the H$_{2}$O maser burst behaviour in S235AB-MIR is caused by the overlapping of maser components along the line of sight, as was seen by \citet{Shimoikura05} in Orion KL, and propose that sub-feature scale structures exist which are not fully resolved at our current angular resolution.

Masers trace a young ($\sim1600 ~\mathrm{years}$) jet which is aligned to outflows seen in HCO$^{+}$ and C$^{34}$S. The maser jet has a position angle of $PA = 217^{\circ}$ and diameter of 15 au, and exhibits a well defined velocity gradient across its axis. The jet has a steep velocity gradient of 1.2 km s$^{-1}$ mas$^{-1}$ which corresponds to 0.77 km s$^{-1}$ au$^{-1}$ at a distance of 1.56 kpc.

3D motions, derived from line-of-sight and sky-plane velocities of 16 water maser features are well described by a rotating cylinder model, giving the water maser jet a manner comparable to a `water spout'. Our best-fit model parameters were $v_{\mathrm{out}} = 45 \pm 2$ km s$^{-1}$, $v_{\mathrm{rot}}=22 \pm 3$ km s$^{-1}$, $i = 12 \pm 2 ^{\circ}$, which are the outflow velocity, tangential rotation velocity and line-of-sight inclination, respectively. Our parameters are consistent with other investigations of jet rotation at similar scales. These new observations support magnetocentrifugal jet rotation as a possible mechanism of dispersing angular momentum from the protostellar disks of accreting massive stars. 


One maser feature was found to accelerate steadily at a rate of $2.92 \pm1.01$ km s$^{-1}$ yr$^{-1}$, which was consistent with that predicted in our rotating jet model. The velocity drift rate is similar to that seen in another DBSM source although likely different in origin.

\null

\section*{Acknowledgments}

We give our sincere gratitude to the anonymous referee who provided a very thorough checking of this manuscript, and whose suggestions lead to many critical improvements to this work.

We would like to thank Luca Moscadelli for sharing ideas and interesting discussion on this maser source.

R.B. would like to acknowledge the Ministry of Education, Culture, Sports, Science and Technology (MEXT), Japan for financial support under the Monbukagakusho scholarship.

H.I. has been supported by Japan Society for the Promotion of Science (JSPS) Grant-in-Aid for Challenging Exploratory Research (25610043).

\def\ref@jnl#1{{\rmfamily #1}}%
\newcommand\aj{\ref@jnl{AJ}}%
\newcommand\araa{\ref@jnl{ARA\&A}}%
\newcommand\apj{\ref@jnl{ApJ}}%
\newcommand\apjl{\ref@jnl{ApJ}}%
\newcommand\apjs{\ref@jnl{ApJS}}%
\newcommand\ao{\ref@jnl{Appl.~Opt.}}%
\newcommand\apss{\ref@jnl{Ap\&SS}}%
\newcommand\aap{\ref@jnl{A\&A}}%
\newcommand\aapr{\ref@jnl{A\&A~Rev.}}%
\newcommand\aaps{\ref@jnl{A\&AS}}%
\newcommand\azh{\ref@jnl{AZh}}%
\newcommand\baas{\ref@jnl{BAAS}}%
\newcommand\jrasc{\ref@jnl{JRASC}}%
\newcommand\memras{\ref@jnl{MmRAS}}%
\newcommand\mnras{\ref@jnl{MNRAS}}%
\newcommand\pra{\ref@jnl{Phys.~Rev.~A}}%
\newcommand\prb{\ref@jnl{Phys.~Rev.~B}}%
\newcommand\prc{\ref@jnl{Phys.~Rev.~C}}%
\newcommand\prd{\ref@jnl{Phys.~Rev.~D}}%
\newcommand\pre{\ref@jnl{Phys.~Rev.~E}}%
\newcommand\prl{\ref@jnl{Phys.~Rev.~Lett.}}%
\newcommand\pasp{\ref@jnl{PASP}}%
\newcommand\pasj{\ref@jnl{PASJ}}%
\newcommand\qjras{\ref@jnl{QJRAS}}%
\newcommand\skytel{\ref@jnl{S\&T}}%
\newcommand\solphys{\ref@jnl{Sol.~Phys.}}%
\newcommand\sovast{\ref@jnl{Soviet~Ast.}}%
\newcommand\ssr{\ref@jnl{Space~Sci.~Rev.}}%
\newcommand\zap{\ref@jnl{ZAp}}%
\newcommand\nat{\ref@jnl{Nature}}%
\newcommand\iaucirc{\ref@jnl{IAU~Circ.}}%
\newcommand\aplett{\ref@jnl{Astrophys.~Lett.}}%
\newcommand\apspr{\ref@jnl{Astrophys.~Space~Phys.~Res.}}%
\newcommand\bain{\ref@jnl{Bull.~Astron.~Inst.~Netherlands}}%
\newcommand\fcp{\ref@jnl{Fund.~Cosmic~Phys.}}%
\newcommand\gca{\ref@jnl{Geochim.~Cosmochim.~Acta}}%
\newcommand\grl{\ref@jnl{Geophys.~Res.~Lett.}}%
\newcommand\jcp{\ref@jnl{J.~Chem.~Phys.}}%
\newcommand\jgr{\ref@jnl{J.~Geophys.~Res.}}%
\newcommand\jqsrt{\ref@jnl{J.~Quant.~Spec.~Radiat.~Transf.}}%
\newcommand\memsai{\ref@jnl{Mem.~Soc.~Astron.~Italiana}}%
\newcommand\nphysa{\ref@jnl{Nucl.~Phys.~A}}%
\newcommand\physrep{\ref@jnl{Phys.~Rep.}}%
\newcommand\physscr{\ref@jnl{Phys.~Scr}}%
\newcommand\planss{\ref@jnl{Planet.~Space~Sci.}}%
\newcommand\procspie{\ref@jnl{Proc.~SPIE}}%

\bibliographystyle{mn2e}
\bibliography{arXiv}

\appendix

\section{Inverse phase referencing technique for VERA data (mode: 7 and 7MM), using POPS} 

As mentioned in Section 2, the inverse phase referencing (IPR) technique was introduced by \citet{Imai12}. Those authors provide a discussion of the advantages of IPR and the challenges of its implementation in dual-beam VLBI. They also provide a link to a webpage hosting working scripts that can run IPR (and conventional phase referencing) written in ParselTongue.
We introduce here an approach to IPR VLBI data reduction using AIPS POPS.

In the IPR approach to reducing VLBI data the phase and rate solutions obtained using the maser emission are used to calibrate the continuum source data. This practice is commonplace for (single beam) VLBI arrays using fast switching and a standardised frequency setup, however it is more complicated using VERA in VERA7 and VERA7MM modes which allocate only a small bandwidth to the maser emission and a much wider bandwidth to the continuum source.

VERA dual-beam observations allow simultaneous observations of maser emission and a continuum source in separate telescope beams, henceforth referred to as the \emph{`A-beam'} and \emph{`B-beam'}, respectively. 

With regards to phase calibration the first step is to apply source specific delay-tracking solutions which are based on tropospheric models calibrated using GPS measurements of the zenith path delay at each station \citep{Honma08b}. Then we normalise phase gradients in the B-beam baselines by solving for the group delay using a bright continuum source.
These solutions are also applied to the maser data since the wide bandwidth of the B-beam provides better determination of phase slope than the narrow bandwidth of the A-beam.  Instrumental phase contributions from the dual-beam system were measured using an artificial noise source and subsequently removed. 

At this stage all a-priori, station position, instrumental and group delay solutions have been applied. What remains are mostly atmospheric phase residuals which, at 22 GHz, are dominated by the dynamic troposphere (\emph{see} \citealt{Asaki07}).

Phase and rate solutions from the maser emission were then obtained with the task FRING in the A-beam at intervals of 1-2 minutes. Solutions were applied to the B-beam data - thus phase-referencing the continuum source.
Finally - the continuum source was imaged, producing a map in which the co-ordinates have an origin at the phase tracking center of the B-beam in RA and Dec - with the continuum source offset somewhat from the map origin.

Since the precise co-ordinates of the continuum source are well-known its actual offset from the phase tracking center should be negligibly small. The measurable offset comes from the solutions carried over from the A-beam; the separation of the maser with respect to the phase tracking center in the A-beam. The offset of the continuum source therefore gives the absolute co-ordinates of the maser.
Fringe-rate solutions from the maser were also applied to the A-beam, thereby self-calibrating it, which produced high quality maser maps and revealed many of the weak maser spots used in this work.

\begin{figure}
\begin{center}
\includegraphics[scale=0.47]{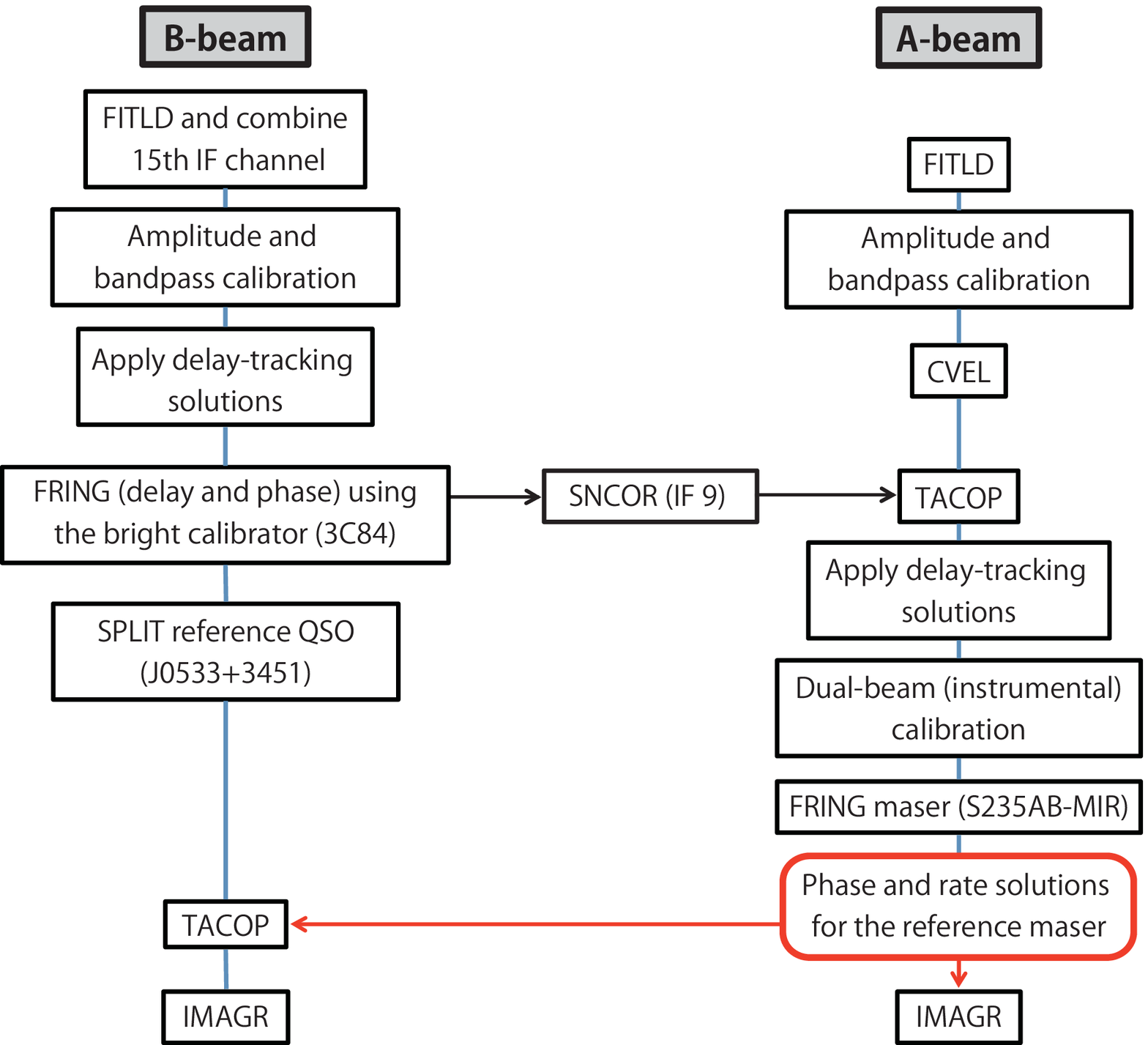}
\caption{Flowchart of the inverse phase-referencing procedure using VERA7/VERA7MM data. \label{FLOW} }
\end{center}
\end{figure}

The flowchart in Figure A1 outlines the reduction method in terms of the main AIPS tasks however a step-by-step reduction guide, in the format of a POPS log file, is hosted at the following webpage:\\

\noindent\footnotesize{\url{http://milkyway.sci.kagoshima-u.ac.jp/~rossburns88/Scripts.html}}

\label{lastpage}

\end{document}